# Generative Design for Performance Enhancement, Weight Reduction, and its Industrial Implications

Bashu Aman*

*Pre-final Year Undergraduate Student*
*Department of Mechanical Engineering, Indian Institute of Technology Kharagpur 721302, West Bengal, India*

**Abstract**

This paper investigates the generative designing of a bracket that aids in the rotation of a linkage mounted on it with a revolute joint. Generative design is a term that is generally used when we care about weight reduction and performance gain in the manufacturing sector. Optimization is also a similar term. The only difference between the two is that generative design considers stress analysis while optimization considers load paths to give the required result. For a critical analysis of a generative designed bracket, Autodesk Fusion 360 is used, which has in-built functionality of working on the generative design on the cloud, given Preserve and Obstacle geometries, Starting shape and Resolution, Load cases and Constraints, Materials and Manufacturing methods. The generative bracket's iterative solutions are explored and further validated with simulations such as stress simulations, thermal simulations, buckling simulations, and modal frequencies, thus leading to a proper result for forming design and mesh. After post-processing the bracket's design and mesh, its CAD is further converted to CAM to send it for 3D printing under a suitable manufacturing method. The final generative bracket design is linked and employed into the initial assembly to check its proper working with the linkage, machine interface, and joints. The generated design considering various parameters is the best finish for performance enhancement and weight reduction. This paper will act as a basis for designing parts involving multiple studies and environments that will be a topic of industrial research in future technologies.

---

\* Author.
*E-mail address:* bashuaman14@iitkgp.ac.in (Bashu Aman).
*Contact*: +91-9262625293
Pre-printed version available from: http://arxiv.org/abs/2007.14138.

# 1. Introduction

Over the past decade, designing has been made easy with the evolution of different tools to help the designers reach their goals. Generative Design, often referred as GD[1] is a tool that uses mathematical formulations and algorithms to optimize a design under particular constraints for minimizing/maximizing an objective function, viz.: Mass, Manufacturing cost, Volume, Maximum von Mises stress, Factor of safety, Maximum displacement [1][2]. The designers explore the suitable outcome from various evolved iterations based on industrial and company's requirements. The process follows a framework that can be categorized into three different stages (as in Figure 1).

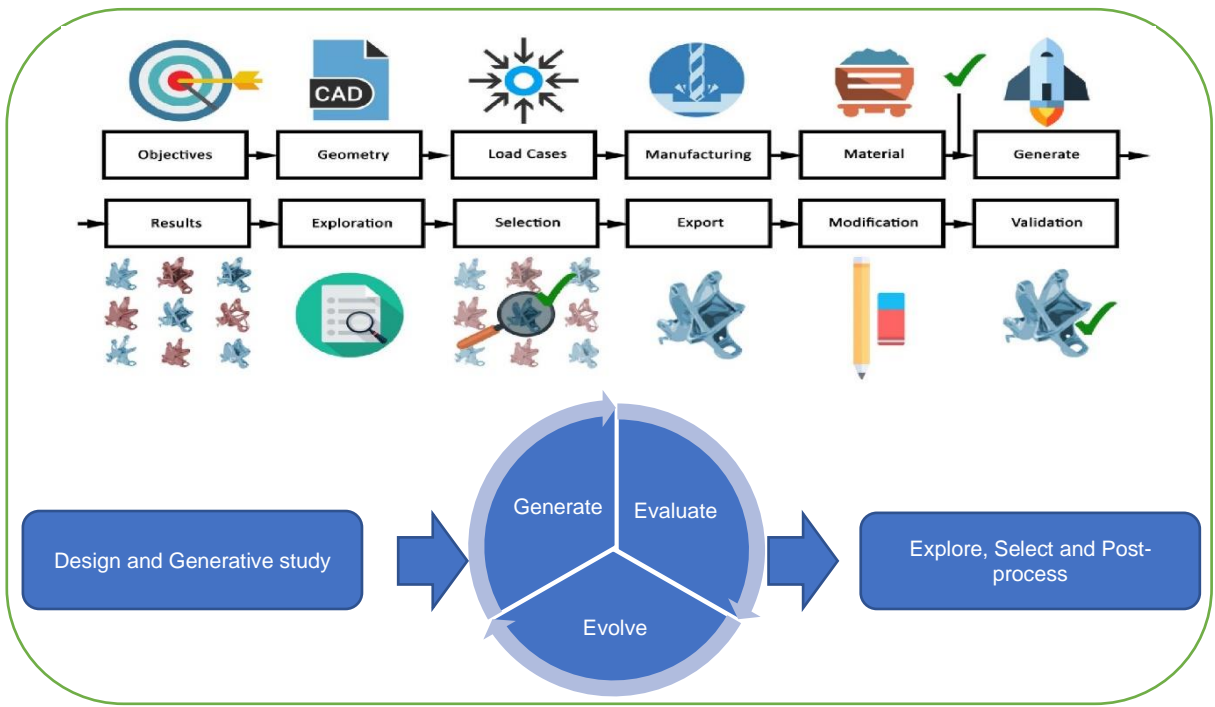

**Figure 1.** GD framework categorized into (a) Pre-GD; (b) GD; (c) Post-GD processes [10]

The pre-GD section involves basic designing, taking into account Preserve and Obstacle geometry, and setting up the study required in the Generative Design workspace by entering parameters for objective, load cases, materials, and manufacturing. GD section involves generating the outcomes by evaluating the study setup and evolving further for a better solution. Later on, post-GD section involves exploration and comparison of the results based on various factors followed by post-processing and validation at the finishing stage. Minimizing the mass is the foremost objective looked onto in the manufacturing sector. Designers try to make the geometry handier and more sophisticated. The design's endurance is tested through a variety of load cases and proper validating simulations. GD tries to convert these requirements into the constraints input during pre-GD and post-GD processes.

GD is possible with various topology optimization software such as Autodesk Fusion 360, and Siemens NX 12 [3]. GD supplements the engineer's imagination. MNC's have also started to bring the change in designs and models by generative design. Volkswagen Innovation and Engineering Centre California retrofitted the generative designed wheel rims, wing-mirror arms, and other parts in a classic 1962 VW Bus [4]. To improve engine performance and reduce vehicle weight, Japan's DESCO Corporation used GD to redesign ECU (Engine Control Unit) that won the iF Design Award for Professional Concept [5]. Technical University of Munich (TUM) in Germany replicated the bones, muscles, and tendons instead of merely putting joints and motors of a Humanoid Robot Roboy 2.0 using GD in order to give humans like finishing and provide stability as well [6]. In San Francisco, Golden Gate Bridge also saw a perfect blend of generative design power when Autodesk Fusion 360 was used to see how the bridge would have looked [7]. Generative design is the future of manufacturing that can surpass today's limitations and provide a design portfolio for innovators. Abode of manufacturing methods find their use in GD, viz.: Unrestricted, Additive manufacturing, or AM, 2.5 axis, 3 axis, 5 axis milling, 2-axis cutting, Die-casting. AM is the most favourable method for enhancing overall performance in terms of lightweight engineering [8].

---

[1] GD stands for Generative Design

Often designers desire a single platform to design, edit, optimize, render, manufacture, validate with simulations, convert CAD to CAM programs and sketch 2D drawings smoothly working with their team. Autodesk Fusion 360 serves the same purpose. Cloud computing is its integral part where all hectic processes can be quickly done on a cloud to overcome unnecessary storage increments and improve the processor's performance. According to Jesse Craft, Jacobs Engineering, NASA's Exploration Portable Life Support System, GD's most exciting and curious thing is that it challenges his biases. He likes right angles, flat surfaces, and round dimensions as an engineer, and according to GD, it may not be the best solution. Hence, if he wants to be the best engineer, he looks to GD to find the optimum solutions [9]. Industries face fierce competition to make their products more efficient, more productive, and everlasting. Consequently, innovators try to come up with more impactful designs for their products that are not only light-weighted but also economically affordable. Generative design learns and evolves from its previous iterations. There is the abode of designs available to choose from during each iteration scaling it on different factors. Convergent status of design is one of the factors that is first and foremost looked onto. It signifies that the design is up to the range of costs, mass, volume, stress, and the factor of safety provided by the designers. Completed status is another factor that signifies of exceeding/ limiting of the design on various aspects. However, designers sometimes find their required design from completed status by looking at the history of the evolved outcomes [10].

Autodesk Fusion 360 has proved its mettle in today's technological era where it can perform millions of permutations in a couple of times that a single human brain cannot even think of doing. Thomas Nagel, Claudius Peters, has used this technology to its full potential as his team reduced the weight of the components, they manufacture 25% less than what was until then thought to be the optimized part [11].

This paper deals with the optimum design required to replace the complex sheet metal lifting bracket design with weldments and four mounting interfaces to the wall that helps the linkage revolve around its mount. It is a more efficient version of sheet metal lifting bracket design as compared to that with two mounting interfaces to the wall (as in Figure 2). In this paper, the complexity (in Figure 3) is reduced by substituting different hardware, viz.: nuts, bolts, screws, and mechanical interfaces with a simple design. Lifting bracket finds it use in several mechanisms involving lifting and transportation jobs as done by cranes, lifts, and automotive factory robots. The bracket finds itself useful in jacking and cornering heavy pieces of equipment and machinery. The bracket is designed, keeping in mind the mass target, the factor of safety limit, and other constraints validated through Finite Element Analysis (FEA). Industries require machinery that has less weight, more efficient, more sophisticated, more promising, and consumes less power. Generative design is the tool that fulfils all these requirements in one go. Hence this paper will provide a base for future enthusiasts who want to learn and excel in this tool.

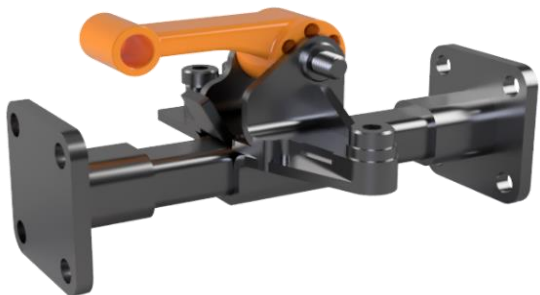

**Figure 2.** Sheet metal lifting bracket with two mounting interfaces to the wall

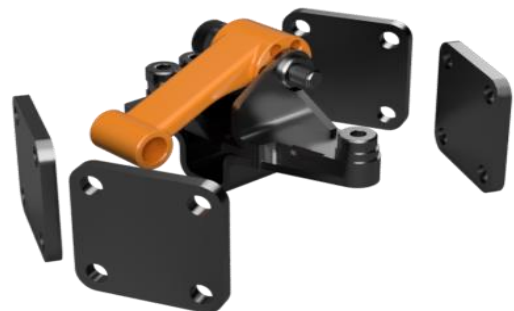

**Figure 3.** Sheet metal lifting bracket with four mounting interfaces to the wall

## 2. Objectives

This paper aims at achieving the undermentioned objectives-

- Getting an optimum bracket design in terms of weight, performance, manufacturing cost, and durability.
- Designing workflow of generative design to innovators.
- Providing a single platform for optimum designing with validating outcomes.
- Setting full dimensions of the latest technology evolved in Autodesk Fusion 360.

## 3. Design and Study Generation

In industries, manufacturing of parts and components of massive pieces of machinery always start with the initial designing. It is the initialization that always matters. With proper initial designing only, the designers can achieve a better generative design solution. Initial designing provides a base for starting topology optimization through mathematical formulations and algorithms. GD is the technology of future engineers [12]. It assists the new generation technologists and is easily compatible. No one can deny this fact. Autodesk Fusion 360 provides a platform for all the generative designing work done on the cloud [10][13]. It also keeps all the enrolled team members updated with the latest up-gradation. New advancements in Autodesk Fusion 360 make it easy for designers to design their initial shape (Also called Starting shape). Hence to start initial designing, knowledge of all the workspaces in Autodesk Fusion 360 is necessary. The available workspaces are-
- Design
- Generative Design
- Render
- Animation
- Simulation
- Manufacture
- Drawing (From Design and From Animation)

These workspaces are linked with each other, and they also provide a different set of tools to achieve the desired target in a particular workspace. Working with designs starts in the Design workspace. Further study setup and generation goes on in the Generative Design workspace. Render, and Animation workspace provides a preview of final design outcomes. The Simulation workspace handles the validation process. The Drawing workspace provides final printing in 2D. Finally, the Manufacture workspace provides manufacturing and CAM conversion [10][13][14].

3.1 *Initialization and Designing*

The generative bracket's proposed design starts initializing, and the process of Preserve and Obstacle geometries designing starts, taking reference from sheet metal lifting bracket design (Figure 3). All the weldments and machine interfaces in sheet metal lifting bracket design

(Figure 3) add to the bracket's mass and reduce performance. The complexity is reduced by generative design mathematical formulations and algorithms to get a design with less weight and more efficiency [2]. The first modifications in the design (in Figure 3) are done either in the Design workspace or Edit model tab in the Generative Design workspace [15]. Both the designing processes have their advantages and disadvantages (Table 1).

**Table 1.** Comparative study between designing in the (a) Design workspace and; (b) Edit model tab in the Generative Design workspace, in Autodesk Fusion 360

| Design workspace | Edit model tab in the Generative Design workspace |
| --- | --- |
| Contains all primary design tabs (Solid, Surface, Sheet metal, Tools) with assemble menu as well. | Contains primary design tabs (Solid, Surface) with no assemble menu. |
| No separate option for New generative model. | Option for New generative model is available. |
| No option for Connector obstacle is present. | Connector obstacle option is available. |
| Any change done in the body or components design will hamper the original design. | Changes in body and components design can be quickly done by duplicating the generative model or using Remove/ Remove all except selected option. |
| Remove features and Remove faces options are not available in the Design workspace. | Options such as Remove features, Remove faces, Replace with primitives are available in the Modify menu. |
| The inspect menu does not have a Transparent surfaces toggle option. | Transparent surfaces toggle option is available in the Inspect menu. |

Initial designing involves designing of Preserve geometry (Geometry needed to be preserved in outcome), Obstacle geometry (Geometry through which outcome should not pass), and Starting shape (If required, geometry, which can be replicated to some extent in outcome). Designing these geometries initializes in the Edit model tab of the Generative Design workspace for simplicity and approachability [15]. The geometries are constructed to setup a generative study in further process. The geometries result in a few of the parameters required in the Generative Design workspace for outcome generation. The mounting holes of the linkage on the bracket and the four machine interfaces with the walls are the required Preserve geometries (as in Figure 4).

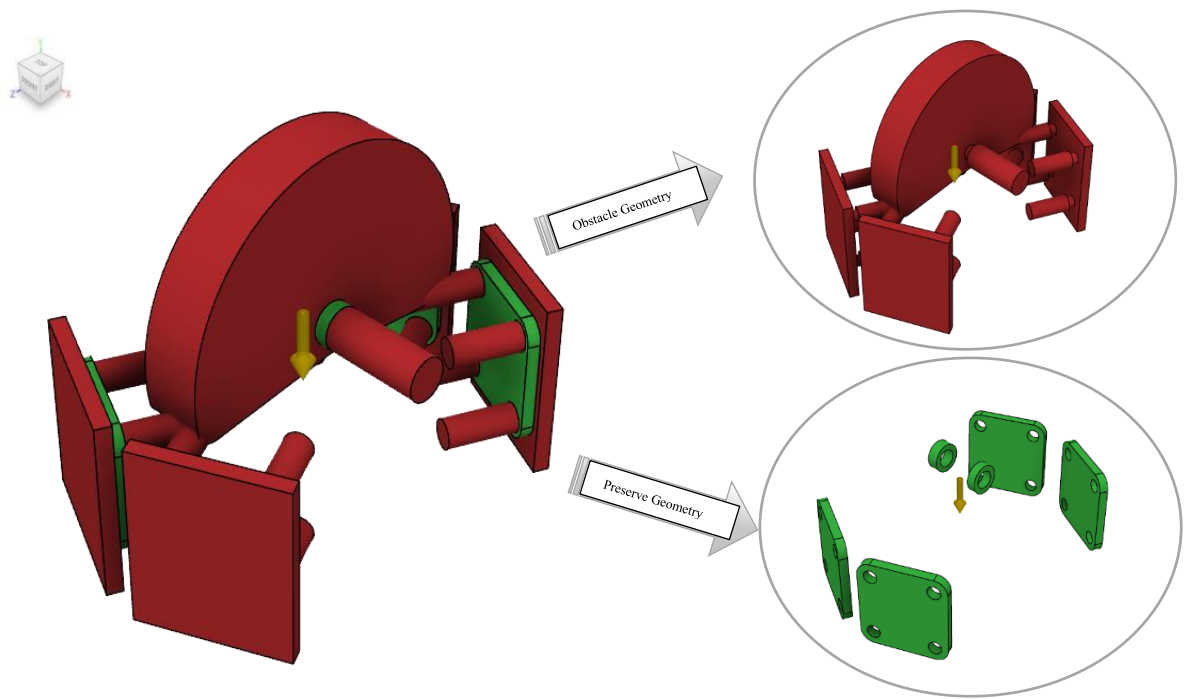

**Figure 4.** (a) Preserve geometries; (b) Obstacle geometries, of the sheet metal lifting bracket as seen in the Generative Design workspace of Autodesk Fusion 360

These Preserve geometries are visualized by the green colour in the Generative Design workspace while setting up the study. The range covered by linkage during its rotation and the bolts required in all the mounting holes are the required Obstacle geometries (as in Figure 4). These Obstacle geometries are visualized by the red colour in the Generative Design workspace while setting up the study. Starting shape is not needed here since there are no additional constraints to limit the bracket's shape. However, if constructed, Starting shape is visualized by the yellow colour in the Generative Design workspace while setting up the study.

3.2 *Generative Model Study Setup*

Study setup begins with the Study settings tab where Synthesis resolution is set somewhat between Coarse and Fine. Synthesis resolution is the indicator of the amount of element size up to which design gets divided for synthesis. More fine means a higher number of elements with a relatively smaller size; thus, it takes more amount of time to generate the outcomes under this resolution. More coarse means a lesser number of elements with a relatively larger size; thus, it takes less amount of time to generate the outcomes under this resolution. The results converge more towards the desired values under Fine resolution than Coarse-resolution.

The study also involves setting Design conditions, which are Structural constraints and Structural loads [10]. The constraints that are structurally applied on the bracket are of utmost importance to the design. Mounting holes cannot be distorted radially due to any manufacturing defect and other forces. Hence, they get fixed along their radial axes. Faces of mounting interfaces with walls also cannot be distorted due to these defects. Hence, all the four faces of the mounting interfaces with the walls get fixed normal to the interfaces. Afterwards, the application of structural loads comes for the analysis of bending, displacement, and the factor of safety. Gravity is the default structural load in all the analysis of designs. In this project, apart from gravity, other structural loads are also taken into consideration. These loads apply to the hinge/ mounting holes of the linkage on the bracket due to forces at different angles applied to the linkage. The loads in a single load case sum up to 500 lb force. The different loads resulting in several load cases are:

- Vertically upward loading of 250 lb force on each mounting hole
- Loading of 250 lb force at 45 deg from vertically upward and radially in the clockwise direction on each mounting hole
- Loading of 250 lb force at 45 deg from vertically upward and radially in the counter-clockwise direction on each mounting hole
- Loading of 250 lb force at 60 deg from vertically upward and radially in the clockwise direction on each mounting hole
- Loading of 250 lb force at 60 deg from vertically upward and radially in the counter-clockwise direction on each mounting hole
- Offset vertically upward loading of 200 lb force and 300 lb force on the two mounting holes and vice-versa
- Off-axis loading of 250 lb force at 5 deg from vertically upward in the clockwise direction on each mounting hole
- Off-axis loading of 250 lb force at 5 deg from vertically upward in the counter-clockwise direction on each mounting hole
- Off-axis loading of 250 lb force at 10 deg from vertically upward in the clockwise direction on each mounting hole
- Off-axis loading of 250 lb force at 10 deg from vertically upward in the counter-clockwise direction on each mounting hole

- Off-axis loading of 250 lb force at 15 deg from vertically upward in the clockwise direction on each mounting hole
- Off-axis loading of 250 lb force at 15 deg from vertically upward in the counter-clockwise direction on each mounting hole

These load cases are applied to provide data points for the bracket's stress analysis in the generative solve. The applied loads may also increase depending on the robust effect of the industrial environments. The final design generates based on all load cases considering material behaviour.

Design criteria also need to be specified for a complete Generative study setup. It provides an option for the design specifications containing Design objectives and Manufacturing methods. These are the parameters related to the goal of the industries they are committing for the design. Design objectives are achieved by either minimizing mass, limiting it to a factor of safety limit, or maximizing stiffness for the desired mass target, limiting it to the factor of safety. For the proposed bracket design, minimizing mass is considered at 4.00 factor of safety, i.e., for a particular amount of load, say 250 lb force, the bracket can sustain 1000 lb force. Today's innovators' main aim is to reduce the weight of the components as much as possible to overcome the difficulties in transportation and handling. Hence this option is a great asset for them. The Design criteria tab also specifies Manufacturing methods. Autodesk Fusion 360 provides a variety of options for manufacturing methods to the designers. The available options are Unrestricted; Additive; 2.5 axis, 3 axis, 5 axis milling; 2-axis cutting, Die-casting [16]. The unrestricted option provides no constraints of manufacturing; it may be a physical constraint or in any other. The additive method is the best possible method in today's industries [8]. Additive manufacturing can be used in the whole manufacturing cycle [17]. Additive manufacturing or AM finds its use in diverse fields such as Biomimicry, Sculpture modelling, Multi-material manufacturing, Model calibration, High precision machines manufacturing, Cellular, and Tubular structure manufacturing [18][19][20][21][22][23]. Today AM is only used to manufacture limited size components. However, it can be used to manufacture large and complex sized components in the industries [24]. Data-driven additive manufacturing is also being considered for improving manufacturing speed and reducing overall time consumption [25]. Milling is also sometimes used depending on the different axes of operation of tools. Milling is an infamous cutting method used in the current industries. Automated robots are used for the 5 axis milling process in the industries [26]. 2-axis cutting method is also quite famous in pipe cutting. These manufacturing methods are multitasked that provide final shape to the components. One can choose more than one option to get a pool of outcomes for better comparison. For the proposed bracket design, Unrestricted; Additive; 3 axis milling with tool direction in X+, X-; 3 axis milling with tool direction in Z+, Z-; 3 axis milling with tool direction in all six directions, viz.: X+, X-, Y+, Y-, Z+, Z-; 5 axis milling method are chosen. Under AM method, Overhang angle and Minimum thickness's specified values are 45 deg and 0.118 in respectively. Under the milling method, specified values of Tool diameter, Tool shoulder-length, Head diameter are 0.375 in, 1.60 in, and 1.50 in, respectively. These values depend on the manufacturing machines used; it may be a milling machine, 3D printing machine, and CNC cutting machine. The choice of design criteria is a significant factor for the optimum design generation. This choice depends on the land, machine, raw materials availability, and other industrial facilities.

Materials also pave the way to minimum weight and maximum performance of the machine's components. Factors such as fabrication of appropriate materials, print materials, and support material volume are always considered for design and structural optimization [22]. Nowadays, digital fabrication considering a massive number of 3D prints as a base in step format using Autodesk Fusion 360, is coming into the picture [27]. Autodesk Fusion 360 provides a bundle of materials libraries. Each library depicts some manufacturing methods, linear materials, non-linear materials, and user favourite materials. The designers can change the properties of the favourite materials. These properties may be the material's identity, appearance, or physical behaviours such as basic thermal properties, mechanical properties, and materials' strength. The choice of material depends on factors varying from industries to industries. Some designers want the material of less mass. Some want the material of less cost to make it economically affordable. Some have the choice of more strong materials in terms of physical behaviour. Some want more thermally resistant material. This option is open for all the designers to select a variety of materials for generative study through generative design. After comparing the results, they can easily choose the best possible option for the respected industries. In this project, the finalized materials for the generative study of the bracket are:

- Aluminium AlSi10Mg (Default option for all the designers)
- Iron, Cast
- Stainless Steel
- Aluminium 5052
- Aluminium 6061
- Aluminium 7075

These materials are selected to cover all the design specifications' primary objectives, i.e., less mass, less cost, more strength, more efficient, readily available, and safe handling. The completed generative setup with all these selections is ready for precheck. Hence the designers can now precheck the setup as it is going for the generation process on the cloud.

3.3 *Preview and Generate Process*

Before the generation process, it is suggested to precheck all the entered values as the missing input may impact the final results. One can preview the sample as well to get an idea of the design outlook. After prechecks, the study is ready for design generation. The generating process is carried on the cloud to save local data storage and processor's life. Every study requires 25 cloud credits to generate outcomes. Cloud credits are currencies used to perform cloud-related jobs such as generation, simulation, and rendering [28]. The generation process is the heart of the generative design.

**4. Outcomes Exploration and Design Finalization**

Since a variety of design criteria and materials were selected earlier for the generative design process. So, many outcomes are available to designers. These outcomes are comparable based on various factors such as Processing status, Study, Visual similarity, Manufacturing method, Material, Piece part cost, Fully burdened cost, Volume, Mass, Maximum von Mises stress, Minimum factor of safety and Maximum displacement global.

4.1 *Exploration*

Differences between the outcome's properties help us to decide the final design. The final design should be the best design among the outcomes. The varied outcomes based on Volume, Material, and Manufacturing method are shown in Table 2.

**Table 2.** Different types of outcomes based on various factors of manufacturing in the Explore tab of Generative Design workspace of Autodesk Fusion 360

| Factors of manufacturing | | Outcomes |
|---|---|---|
| Volume | | 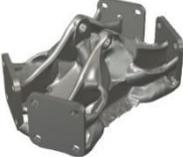 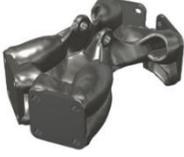 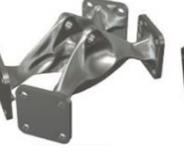 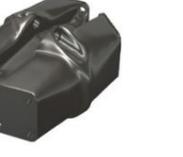 |
| Material | | 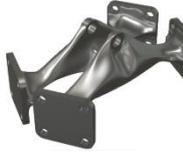 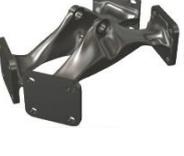 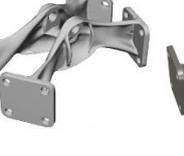 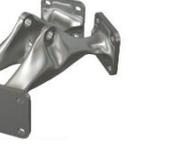 |
| Manufacturing method | Unrestricted | 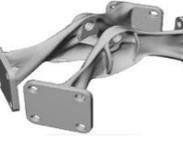 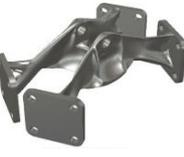 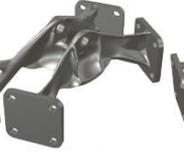 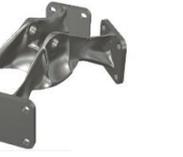 |
| | Additive | 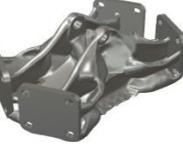 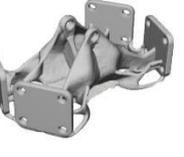 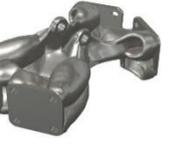 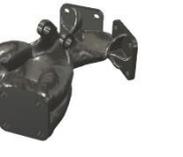 |
| | 3 axis milling | 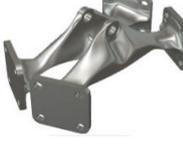 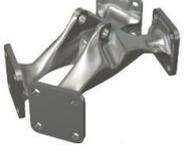 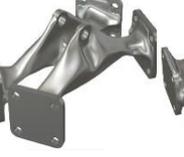 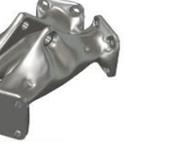 |
| | 5 axis milling | 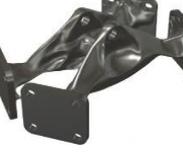 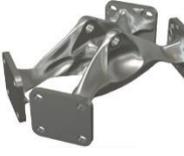 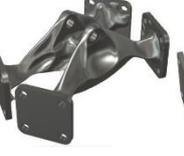 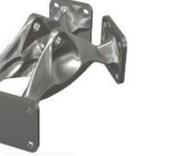 |

Comparing the properties of the different outcomes is also necessary for achieving the desired target. The scatter plot graphs in Figure 5 and Figure 6 compares the manufacturing methods and materials used to generate the outcomes respectively based on Mass, Fully burdened cost, Maximum von Mises stress, Minimum factor of safety, and Maximum displacement global. Here, these quantitative parameters provide a reference for selecting the best design. The best outcomes which suit the project's objective are selected.

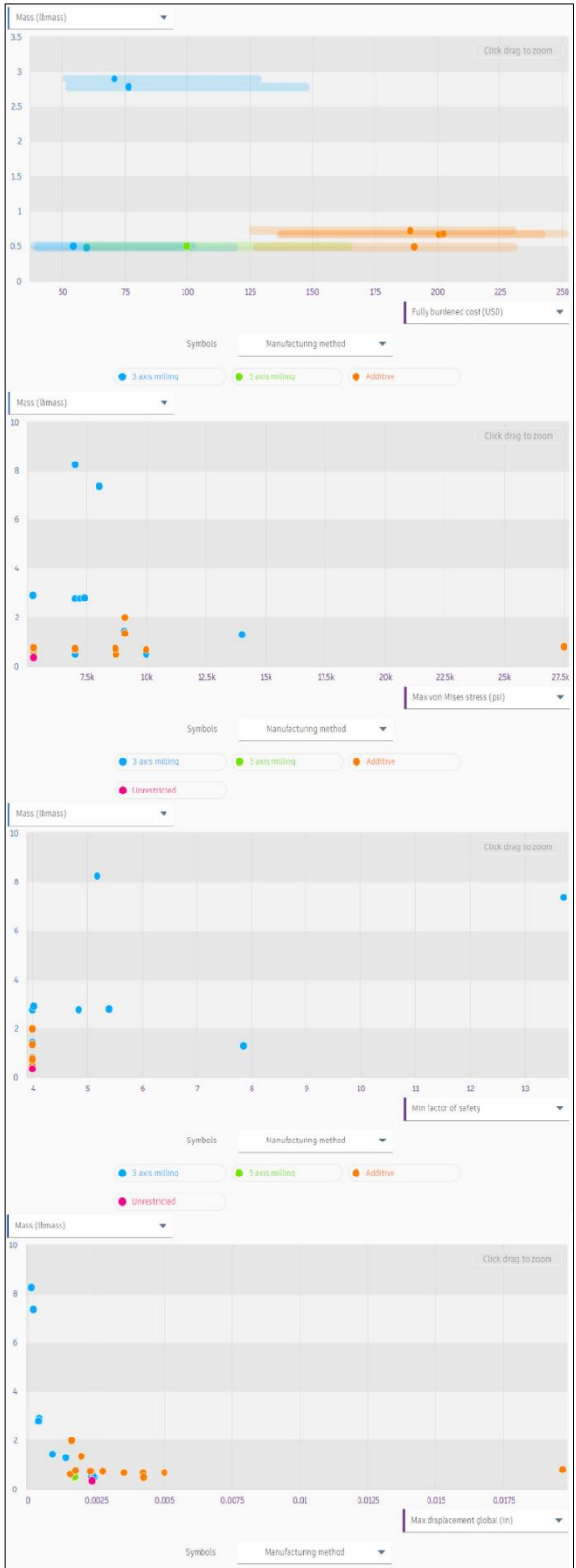

**Figure 5.** Comparing the manufacturing methods that are used to generate the outcomes based on Mass and (a) Fully burdened cost; (b) Maximum von Mises stress; (c) Minimum factor of safety; (d) Maximum displacement global

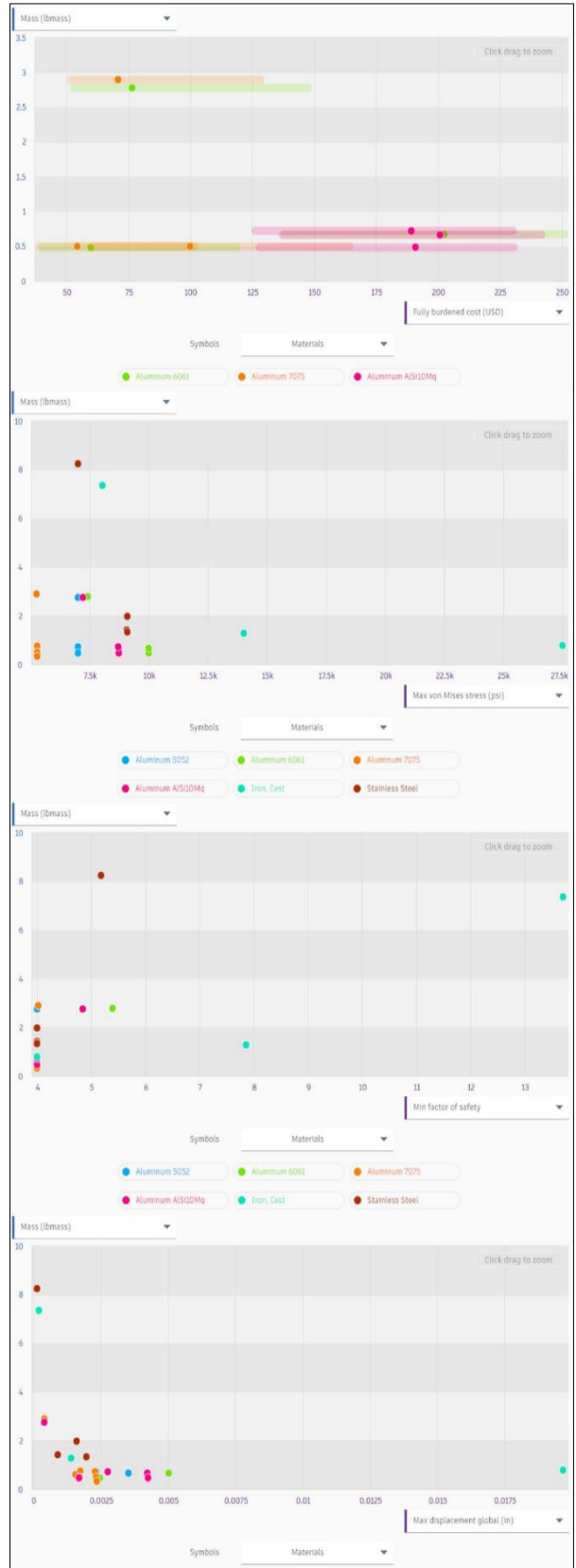

**Figure 6.** Comparing the materials that are used to generate the outcomes based on Mass and (a) Fully burdened cost; (b) Maximum von Mises stress; (c) Minimum factor of safety; (d) Maximum displacement global

Table 3 compares the four best and suitable outcomes to achieve the design objective.

**Table 3.** Comparison between the four best and suitable outcomes from a variety of outcomes in the Explore tab of Generative Design workspace of Autodesk Fusion 360, to achieve the project's main objective

| Properties | Outcome 1 | Outcome 2 | Outcome 3 | Outcome 4 |
|---|---|---|---|---|
| Design preview | 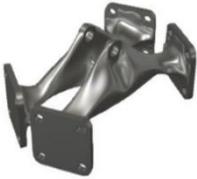 | 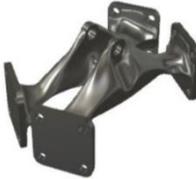 | 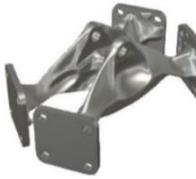 | 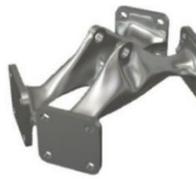 |
| Stress view | 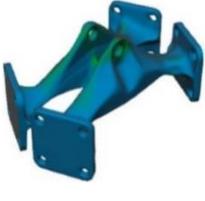 | 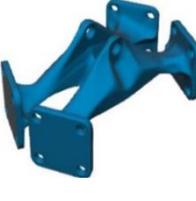 | 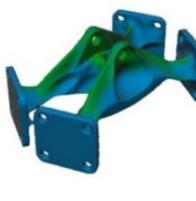 | 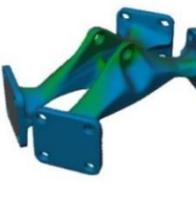 |
| Material | Stainless steel | Iron, cast | Aluminum 7075 | Aluminum 5052 |
| Orientation | X+,Y+,Z+,X-,Y-,Z- | X+,Y+,Z+,X-,Y-,Z- | Unrestricted | X+,Y+,Z+,X-,Y-,Z- |
| Manufacturing method | 3 axis milling | 3 axis milling | 5 axis milling | 3 axis milling |
| Mass (lb mass) | 1.423 | 1.277 | 0.496 | 0.476 |
| Volume (in$^3$) | 4.93 | 4.94 | 4.89 | 4.92 |
| Max. von Mises stress (psi) | 9057.5 | 13989.1 | 5255.3 | 6995.2 |
| Min. factor of safety | 4 | 7.86 | 4 | 4 |
| Max. displacement global (in) | 8.819e-4 | 0 | 0 | 0 |

## 4.2 Design Finalization

After analyzing the scatter plot graphs in Figure 6, Aluminium 5052 has comparatively less mass than other materials. It has a max. von Mises stress lower than 7.5k psi. The min. factor of safety is 4, which is well within the given limit. The max. displacement global is also lower than 0.005 in. The scatter plot graphs in Figure 5 show that 3 axis milling has the least fully burdened cost, lying in the range of 0-120 USD for the mass range of Aluminium 5052. Also, 3 axis milling is the most suitable manufacturing method, which lies approximately in the same range as Aluminium 5052, considering other parameters.

After comparing the best feasible outcomes in Table 3, outcome 4 comes out to be the optimum solution for the generative design. It meets the design objective of weight reduction and performance enhancement that is the need of the hour in the industries. Its manufacturing method is 3 axis milling, and the material is Aluminium 5052, which has the best design specifications. The optimum solution also reduces the extra mass of the weldments and additional hardware. Preserve, and Obstacle geometries are taken care of in the final design (Figure 7). This optimum outcome is finally converted to design form and mesh form (Figure 8). Both the forms provide us a look at the final design before rendering. It also helps the designers to remove any unwanted elements in the design. T-splines can also be edited. The form can be made smooth. These steps take place in the post- processing of the design and mesh.

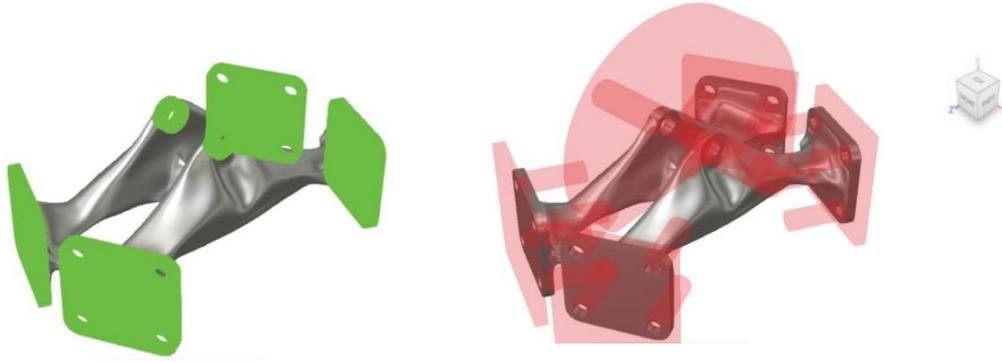

**Figure 7.** Optimum design of the bracket generated, with (a) Preserve geometries (left); (b) Obstacle geometries (right)

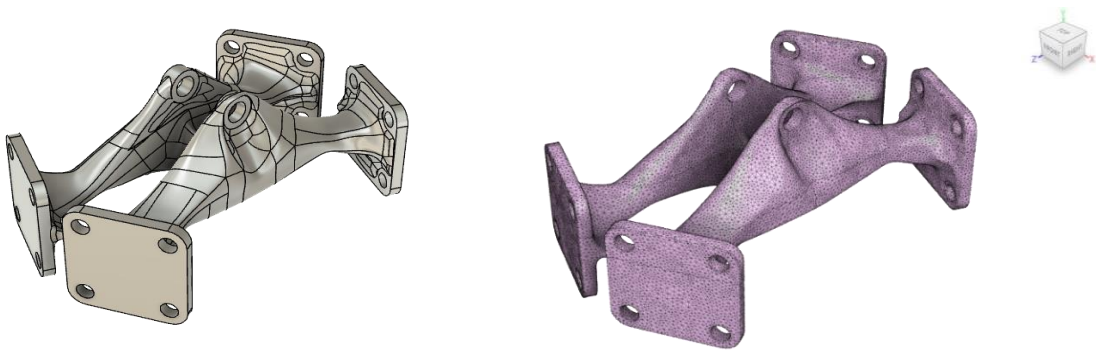

**Figure 8.** (a) Design form (left); (b) Mesh form (right), of the generative designed bracket before post-processing

## 5. Post-Process and Validation

Generative designed components imported from outcomes may have some irregularities. There may be few areas of defects in it. Hence, post-processing the design and mesh form is necessary to achieve proper validation of the final output. Afterwards, validation with the help of simulations checks the design's stability in real case scenarios.

### 5.1 *Post-Processing GD Bracket*

Post-processing is useful for better surface finish and improvement of surface roughness. It adds more functionality to the 3D printed structures. It also improves the fatigue strength of the manufactured structural parts [29]. It can be achieved in many ways, from the coating, polishing, UV curing, heating to ultrasonic abrasion finishing, vibratory bowl abrasion, and laser ablation [30][31]. There are numerous operations available in the Design workspace of Autodesk Fusion 360 for post-processing. The symmetrical body from the GD outcome can be created easily using the midplane feature [32]. T-splines can be edited and reformed as per the designer's wish [33]. GD outcomes can also be smoothed to a certain level. Push/ Pull option is available to the designers for increasing or decreasing the amount of the extrude. A variety of modification tools can be accessed from the Modify tab. Some of them are Insert edge, Subdivide, Insert point, Merge edge, Bridge, Fill hole, Erase and Fill, Weld vertices, UnWeld edges, Crease, UnCrease, Bevel edge, Slide edge, Smooth, Pull, Flatten, Straighten, Match, Interpolate, Thicken, Freeze. Different display modes are available for carrying out the modifications, i.e., Box display, Control frame display, Smooth display.

There are some irregularities in the generative designed bracket. Table 4 shows the GD bracket's post-processing by removing unwanted and excess materials near the adjoining machine interfaces that mount the bracket to the wall. These are low-stress regions. Hence this step leads to an ideal stressed structure. When the bracket assemblies with the linkage, there is a high-stress region also. It is present on the bracket at a location below the linkage. Hence in the post-process, more material gets added, and the resulting surface smoothed to remove this defect. Also, the linkage mount is extra extruded (as in Figure 9) for the machine's proper handling. All significant defects get removed from the GD bracket. Hence the bracket is post-processed and ready for validation with simulations.

**Table 4.** Comparison of the machine interface region of the generated bracket that mounts the bracket to the wall (a) before and; (b) after post-processing

| Before post-processing the generative designed bracket | After post-processing the generative designed bracket |
|---|---|
| 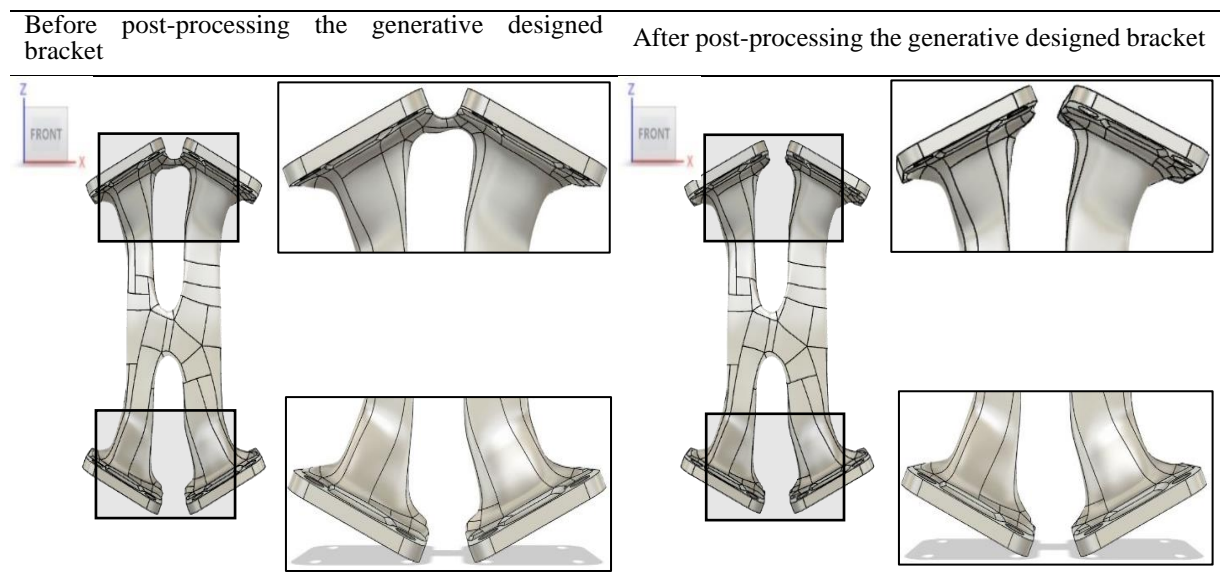 | |

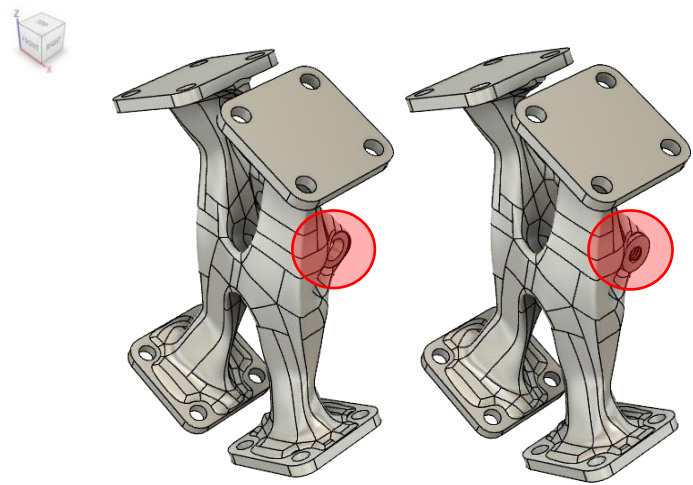

**Figure 9.** Linkage mount of the generative designed bracket (a) before extra extruding (left); (b) after extra extruding (right), during post-processing

*5.2 Validation*

Every processed design component requires validation with the help of simulations so that its design can be modified. Simulations are necessary for predicting the life of the component. Simulations have achieved significant milestones in today's industry and will be an excellent option for validating emerging technological tools in the future industries also [34]. Autodesk Fusion 360 provides a variety of simulations for the designers to validate and optimize their design specifications. They are Static stress, Modal frequencies, Electronics cooling (Preview), Thermal, Thermal stress, Structural buckling, Nonlinear static stress, Event simulation, and Shape optimization. These different simulations help the designers achieve a proper post-processed design based on different factors such as stress, vibration, heat, and buckling. Buckling is a failure criterion that arises by the bending of parts due to compression or elongation due to structural loads applied. Buckling can be considered in a wide variety of areas, i.e., Plastic buckling, Force limiting buckling, Laterally constrained rods, Cellular columns, Long plates, Stiffened panels, and Carbon nanotubes [35][36][37][38][39][40][41][42]. Modal frequency analysis validates a structure by vibrational characteristics arising from dynamic loading on the mechanical structure [43]. Simulation can also be extended to forming processes. Sheet metal forming simulation indicates a broader dimension of simulation assistance in manufacturing processes [44]. Optimization based on hybrid simulation is in advancement in today's technological world where simultaneous consideration is taken between the complex system of material flows and the thermal-physical behaviour of the manufacturing processes to make products more efficient and productive [45][46].

In the proposed project, the Simulation workspace of Autodesk Fusion 360 validates the post-processed GD bracket. Setup for Simulation study gets fed from the generative design itself. All the design conditions, and criteria, viz. Structural constraints, Structural loads, and material of the final design get added from the GD setup itself. Contact tolerance is 0.1 mm, and Model-based mesh size is 10%. All contact sets and overlapping bodies become contacted through the Automatic contacts option. Unnecessary features such as mounting holes thread that affect the simulation result are simplified. After proper pre-check, the simulation model starts solving. It requires 5 cloud credits to solve the simulation model. One can explore the simulation results through animation, comparison, inspection through probes, and slice planes after the simulation model gets solved. The display is toggled to Wireframe visibility.

The detailed report for the generative designed bracket's simulation results is printed in the HTML webpage format and is available from Mendeley Data [47].

For reference, material properties of Aluminium 5052 used in the final GD bracket are shown in Figure 10. Result summary for the simulation of GD bracket under applied Design conditions, Design criteria and selected material is shown in Figure 11. The best simulation output resulting from the Vertical loading load case are shown in Figure 12, Figure 13 and, Figure 14. Hence the optimum post-processed GD bracket is the best solution the designers can get. It is validated by simulation results also. Bracket has not reached its fracture limit under given Design criteria, Design conditions and selected materials. One can convert this design's CAD to CAM in the Manufacturing workspace of Autodesk Fusion 360, which one can further feed into the milling machine. The proposed design aligns to the objective of weight reduction and performance enhancement which is the goal of the innovators in today's industries.

| Density | 2.68E-06 kg / mm^3 |
| --- | --- |
| Young's Modulus | 70300 MPa |
| Poisson's Ratio | 0.33 |
| Yield Strength | 193 MPa |
| Ultimate Tensile Strength | 228 MPa |
| Thermal Conductivity | 0.138 W / (mm C) |
| Thermal Expansion Coefficient | 2.38E-05 / C |
| Specific Heat | 880 J / (kg C) |

**Figure 10.** Material properties of Aluminium 5052 used in the generative design process

| Name | Minimum | Maximum |
| --- | --- | --- |
| **Stress** | | |
| Von Mises | 0.001401 MPa | 56.27 MPa |
| 1st Principal | -5.717 MPa | 60.64 MPa |
| 3rd Principal | -39.66 MPa | 15.16 MPa |
| Normal XX | -33.5 MPa | 56.55 MPa |
| Normal YY | -18.46 MPa | 35.31 MPa |
| Normal ZZ | -28.08 MPa | 45.35 MPa |
| Shear XY | -20.7 MPa | 20.83 MPa |
| Shear YZ | -21.97 MPa | 24.84 MPa |
| Shear ZX | -14.36 MPa | 15.19 MPa |
| **Displacement** | | |
| Total | 0 mm | 0.05809 mm |
| X | -0.05029 mm | 0.04173 mm |
| Y | -1.815E-04 mm | 0.03033 mm |
| Z | -0.01451 mm | 0.01355 mm |
| **Reaction Force** | | |
| Total | 0 N | 52.56 N |
| X | -26.98 N | 24.97 N |
| Y | -28.88 N | 5.43 N |
| Z | -50.16 N | 52.56 N |
| **Strain** | | |
| Equivalent | 3.691E-08 | 0.001142 |
| 1st Principal | -1.672E-07 | 0.001178 |
| 3rd Principal | -7.82E-04 | 6.125E-08 |
| Normal XX | -4.706E-04 | 7.678E-04 |
| Normal YY | -3.683E-04 | 4.941E-04 |
| Normal ZZ | -3.823E-04 | 6.328E-04 |
| Shear XY | -7.833E-04 | 7.881E-04 |
| Shear YZ | -8.315E-04 | 9.4E-04 |
| Shear ZX | -5.432E-04 | 5.749E-04 |

**Figure 11.** Result summary of the simulation report of GD bracket under applied Design conditions, Design criteria, and selected material

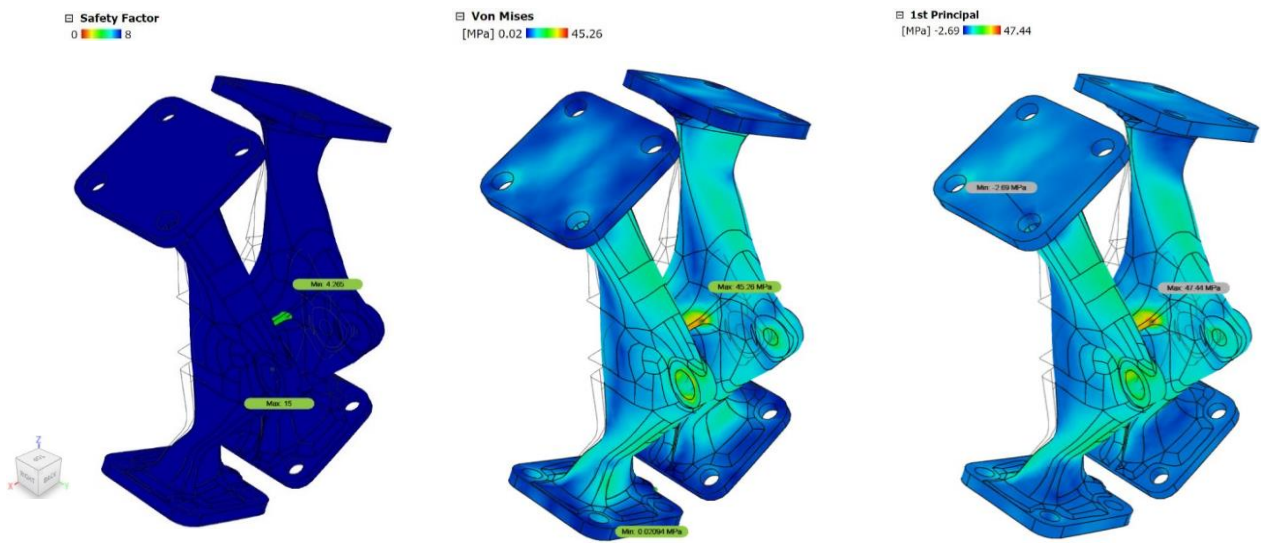

**Figure 12.** Simulation output from Vertical loading load case, in terms of (a) Safety Factor; (b) Von Mises; (c) 1st Principal Stress

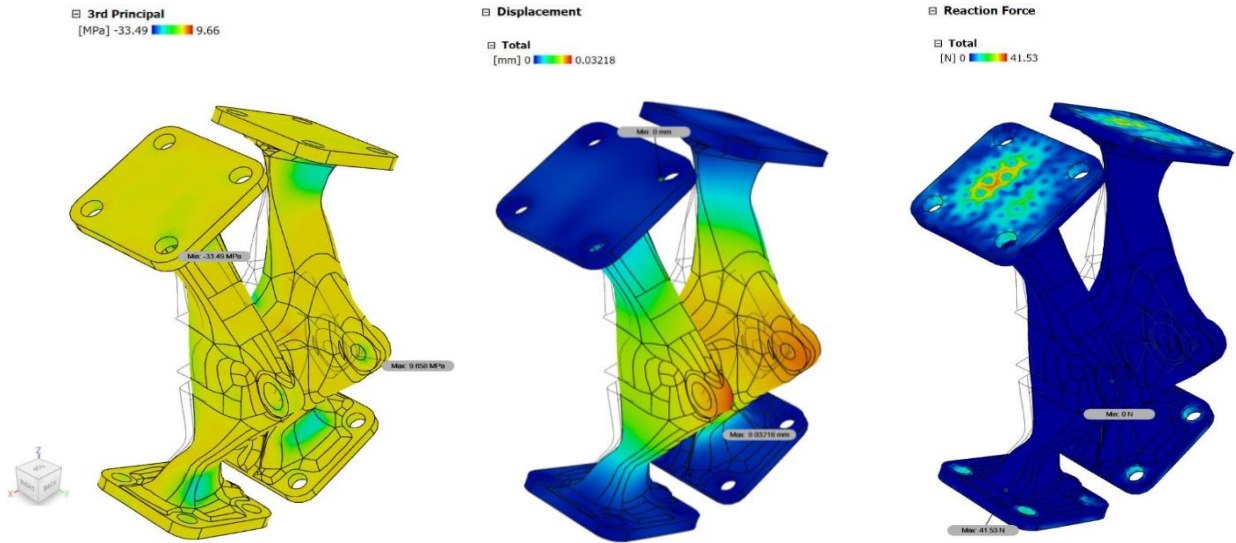

**Figure 13.** Simulation output from Vertical loading load case, in terms of (a) 3rd Principal Stress; (b) Total Displacement; (c) Total Reaction Force

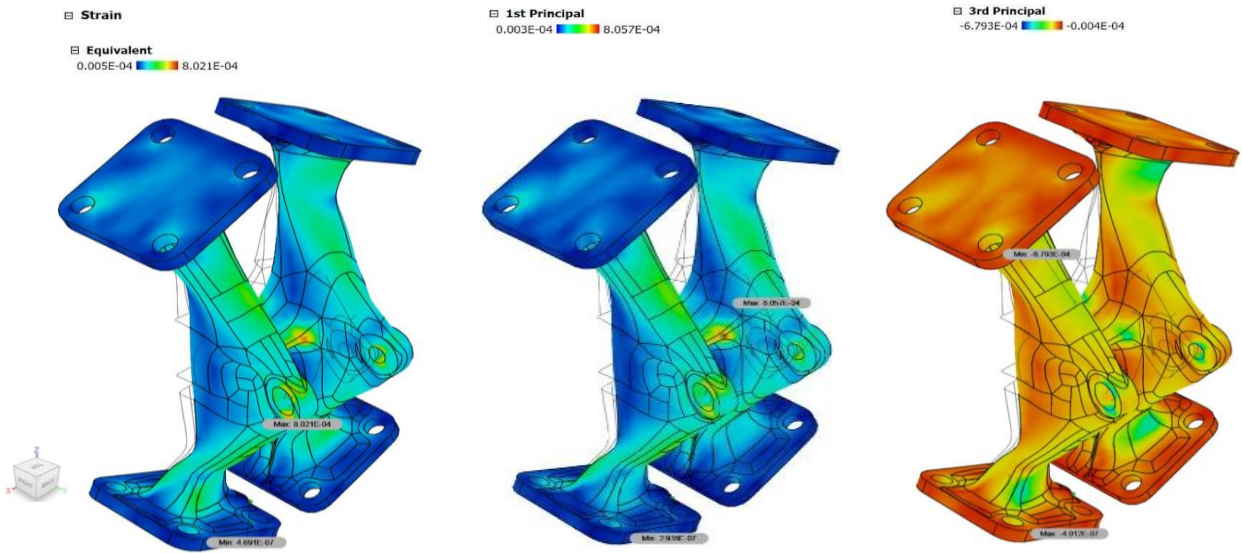

**Figure 14.** Simulation output from Vertical loading load case, in terms of (a) Equivalent Strain; (b) 1st Principal Strain; (c) 3rd Principal Strain

## 6. Conclusions

The proposed work is aimed at achieving an optimum bracket design under the given design specifications and applied environments using generative design as the tool in Autodesk Fusion 360. The post-processed and validated GD bracket, when inserted and linked to the initial assembly (as in Figure 15), enhances the linkage performance, reduces the overall weight and aids in the unhindered rotation of the linkage without any distortion. The GD bracket is made up of Aluminium 5052, the factor of safety of the GD bracket is within the specified limit, and it has not reached its fracture limit in, the simulation process. Hence the manufactured bracket is affordable and durable. This paper dictates the design and validation workflow of generative design to the innovators in the industries. The proposed work also provides a single platform for optimum designing of a wide range of machines through critical analysis of generated and validated outcomes. GD tool has a vast scope of improvement in the manufacturing sector and is a hot topic for industrial research. Autodesk Fusion 360 has done a tremendous job in advancing generative design technology to this extent. However, the generative design has a high probability of being autonomous, bio-inspired and bringing industrial revolution to the printers [48][49][50][51]. In all, it concludes that generative design is a tool that can revolutionize the future manufacturing industries and provide a basis for designing to the industry experts soon.

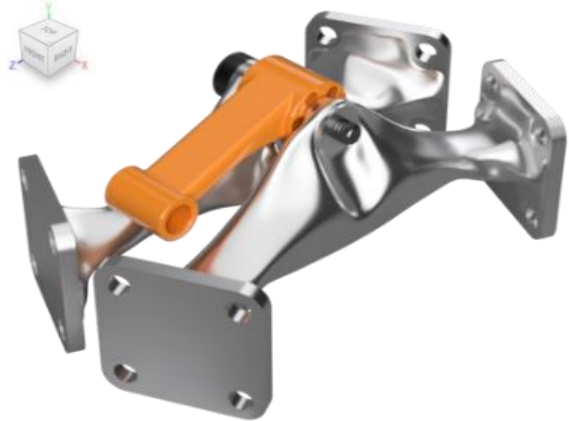

**Figure 15.** The final assembly of the GD bracket and the linkage

### Acknowledgements

The author would like to acknowledge the "Generative Design for Manufacturing Specialization" by Autodesk on Coursera for providing knowledge about the article background, and Indian Institute of Technology Kharagpur for providing free licenses of Coursera and Autodesk Fusion 360. The author would also like to acknowledge Dr. Cheruvu Siva Kumar, Professor, Department of Mechanical Engineering, IIT Kharagpur for his timely scholarly advice, meticulous scrutiny and scientific approach.

### Funding Information

The author received free license of courses on Coursera from Indian Institute of Technology Kharagpur as a part of its COVID initiative. The author also received free student license of Autodesk Fusion 360 from Indian Institute of Technology Kharagpur for preparation of the article.

# Figure Captions

**Figure 1.** GD framework categorized into (a) Pre-GD; (b) GD; (c) Post-GD processes [10]
**Figure 2.** Sheet metal lifting bracket with two mounting interfaces to the wall
**Figure 3.** Sheet metal lifting bracket with four mounting interfaces to the wall
**Figure 4.** (a) Preserve geometries; (b) Obstacle geometries, of the sheet metal lifting bracket as seen in the Generative Design workspace of Autodesk Fusion 360
**Figure 5.** Comparing the materials that are used to generate the outcomes based on Mass and (a) Fully burdened cost; (b) Maximum von Mises stress; (c) Minimum factor of safety; (d) Maximum displacement global
**Figure 6.** Comparing the manufacturing methods that are used to generate the outcomes based on Mass and (a) Fully burdened cost; (b) Maximum von Mises stress; (c) Minimum factor of safety; (d) Maximum displacement global
**Figure 7.** Optimum design of the bracket generated, with (a) Preserve geometries (left); (b) Obstacle geometries (right)
**Figure 8.** (a) Design form (left); (b) Mesh form (right), of the generative designed bracket before post-processing
**Figure 9.** Linkage mount of the generative designed bracket (a) before extra extruding (left); (b) after extra extruding (right), during post-processing
**Figure 10.** Material properties of Aluminium 5052 used in the generative design process
**Figure 11.** Result summary of the simulation report of GD bracket under applied Design conditions, Design criteria, and selected material
**Figure 12.** Simulation output from Vertical loading load case, in terms of (a) Safety Factor; (b) Von Mises; (c) 1st Principal Stress
**Figure 13.** Simulation output from Vertical loading load case, in terms of (a) 3rd Principal Stress; (b) Total Displacement; (c) Total Reaction Force
**Figure 14.** Simulation output from Vertical loading load case, in terms of (a) Equivalent Strain; (b) 1st Principal Strain; (c) 3rd Principal Strain
**Figure 15.** The final assembly of the GD bracket and the linkage

# Table Captions

**Table 1.** Comparative study between designing in the (a) Design workspace and; (b) Edit model tab in the Generative Design workspace, in Autodesk Fusion 360
**Table 2.** Different types of outcomes based on various factors of manufacturing in the Explore tab of Generative Design workspace of Autodesk Fusion 360
**Table 3.** Comparison between the four best and suitable outcomes from a variety of outcomes in the Explore tab of Generative Design workspace of Autodesk Fusion 360, to achieve the project's main objective
**Table 4.** Comparison of the machine interface region of the generated bracket that mounts the bracket to the wall (a) before and; (b) after post-processing